\documentclass[aps,pra,twocolumn,showpacs,amsmath,amssymb,preprintnumbers,superscriptaddress,10pt]{revtex4-1}

\usepackage{bm}
\usepackage{graphicx}
\graphicspath{{figures/}}
\usepackage{SIunits}
\usepackage{hyphenat}
\usepackage{scrextend}
\usepackage[bookmarks=false]{hyperref}
\usepackage{color}
\usepackage{makecell}
\usepackage[capitalise]{cleveref}
\crefname{section}{Sec.}{Secs.}
\Crefname{section}{Section}{Sections}

\def\ket#1{\left| #1 \right\rangle}

\definecolor{pink}{RGB}{255,0,255}

\definecolor{red}{rgb}{1,0,0}

\definecolor{blue}{rgb}{0,0,1}

\begin{document}
\title{Quantum key distribution with distinguishable decoy states}

\author{Anqi~Huang}
\email{angelhuang.hn@gmail.com}
\affiliation{Institute for Quantum Information \& State Key Laboratory of High Performance Computing, College of Computer, National University of Defense Technology, Changsha 410073, People's Republic of China}
\affiliation{Institute for Quantum Computing, University of Waterloo, Waterloo, ON, N2L~3G1 Canada}
\affiliation{\mbox{Department of Electrical and Computer Engineering, University of Waterloo, Waterloo, ON, N2L~3G1 Canada}}

\author{Shi-Hai~Sun}
\email{shsun@nudt.edu.cn}
\affiliation{College of Science, National University of Defense Technology, Changsha 410073, People's Republic of China}

\author{Zhihong~Liu}
\affiliation{College of Mechatronic Engineering and Automation, National University of Defense Technology, Changsha 410073, People's Republic of China}

\author{Vadim~Makarov}
\affiliation{Russian Quantum Center and MISIS University, Moscow}
\affiliation{Department of Physics and Astronomy, University of Waterloo, Waterloo, ON, N2L~3G1 Canada}


\begin{abstract}
The decoy state protocol has been considered to be one of the most important methods to protect the security of quantum key distribution (QKD) with a weak coherent source. Here we test two experimental approaches to generating the decoy states with different intensities: modulation of the pump current of a semiconductor laser diode, and external modulation by an optical intensity modulator. The former approach shows a side-channel in the time domain that allows an attacker to distinguish s signal state from a decoy state, breaking a basic assumption in the protocol. We model a photon-number-splitting attack based on our experimental data, and show that it compromises the system's security. Then, based on the work of K.~Tamaki {\it et al.}\ [New J. Phys. {\bf 18}, 065008 (2016)], we obtain two analytical formulas to estimate the yield and error rate of single-photon pulses when the signal and decoy states are distinguishable. The distinguishability reduces the secure key rate below that of a perfect decoy-state protocol. To mitigate this reduction, we propose to calibrate the transmittance of the receiver (Bob's) unit. We apply our method to three QKD systems and estimate their secure key rates.
\end{abstract}


\maketitle

\section{Introduction}

Unconditional security, or information theoretical security, is always the end goal of cryptography. It is a challenging task for classical cryptography, which is based on mathematical complexity. Luckily, quantum cryptography, based on the basic principle of quantum mechanics, provides a way to reach such terminal goal. Taking advantage of conventional techniques from optical communication and cutting-edge technologies from quantum optics, some quantum cryptography primitives, such as quantum key distribution (QKD)~\cite{bennett1984}, quantum coin tossing (QCT)~\cite{bennett1984}, quantum digital signature (QDS)~\cite{gottesman2001} and quantum secret sharing (QSS)~\cite{cleve1999}, have been implemented~\cite{schmitt-manderbach2007,rubenok2013,pappa2014,yin2017a,collins2016,bogdanski2008}. Of these, QKD implementations are relatively mature, have been demonstrated in a field environment~\cite{stucki2011, sasaki2011, tang2014a,yin2016} and commercialized~\cite{comqkdsystems}.

However, along with QKD maturation, deviation of the behaviour in implementation from the theoretical assumptions has been explored. Such deviation might open backdoors and be exploited by an eavesdropper (Eve). Once any security assumptions are broken by Eve, she could compromise theory-proved security and spy on the secret key in practice. The feasibility and capability of Eve's attacks have been shown in various cases~\cite{xu2010,lydersen2010a,gerhardt2011,sun2011,jain2011,tang2013,bugge2014,sajeed2015a,huang2016,makarov2016}. Similar vulnerability may occur in the other quantum cryptography implementations~\cite{yin2017a, collins2016, bogdanski2008}, since they employ similar optical components and devices.

In practical quantum cryptography systems, a weak coherent source (WCS) is widely used to replace the single photon source. An inherent imperfection in WCS is the emission of multiphoton pulses, which gives Eve more than one copy of Alice's quantum states. Then Eve could perform the photon-number-splitting (PNS) attack~\cite{huttner1995,brassard2000}, in which she blocks all single-photon pulses, and keeps one photon from the multiphoton state. Then she could get the entire final key after Alice and Bob announce their basis choices. Note that a modified PNS attack based on a beam splitter has been demonstrated~\cite{liu2011a}. Thus, the danger of PNS attack is not only theoretical, but also practical. Fortunately, decoy state protocols~\cite{hwang2003, wang2005a, lo2005} were proposed to beat such attack, and have been implemented in many QKD systems~\cite{schmitt-manderbach2007, rosenberg2007, liu2010, tang2014a, yin2016}. They have also been employed in other quantum cryptography systems~\cite{bogdanski2008, fu2015, yin2017a, yin2017} to guarantee their security.

Generally speaking, in the decoy state protocol, signal and decoy states only have different mean photon numbers. Decoy states are used to estimate the detection gain and error rate of single-photon pulses in signal states. If Eve could not distinguish the signal and decoy states, she would change the photon number in both signal and decoy pulses during the PNS attack~\cite{huttner1995,brassard2000}. Thus, she would disturb the yield and error rate of decoy states, which affects the estimation of single-photon detection gain and error rate in signal states. It results in a decrease of the secure key rate~\cite{lo2005}.

However, the essential assumption, indistinguishability of the signal and decoy states, may not be guaranteed in practice. In fact, Eve might exploit practical imperfections to find a side-channel which allows her to distinguish the signal and decoy states. Then she could perform different hacking strategies to keep the normal statistic distributions, meanwhile spy some secret information silently without being discovered. Several types of source imperfections and corresponding attacks have been shown in different QKD systems~\cite{wang2009, nauerth2009, jiang2012, tang2013, tamaki2016}. Importantly, the first quantum satellite also employs one of such imperfect sources~\cite{liao2017}. Recently, the security of decoy-state QKD with a leaky source was proved by K.~Tamaki {\it et al.}\ in Ref.~\onlinecite{tamaki2016}. According to their analysis, the trace distance can be used to characterize the leaked information due to the imperfection of source, which may come from Alice's imperfect state preparation or Eve's attack (such as the Trojan-horse attack~\cite{vakhitov2001,gisin2006,jain2014}). Then the key rate can be bounded with the trace distance, in which the yield and error rate of single-photon pulses are estimated by solving an optimization problem. In this Article, based on the work of Ref.~\onlinecite{tamaki2016}, we obtain two analytical formulas to estimate the yield and error rate of the single photon pulses, which produce the secure key rate given that the signal and decoy states are distinguishable. A further analysis shows that the key rate can be improved by calibrating the transmittance of Bob's optical devices (the calibrated method has also been used to protect a single photon detector of Bob from the blinding attack~\cite{maroy2010,maroy2017}).

In this Article, we first recap the basic theory of decoy state protocol in~\cref{basic}. In~\cref{measurement}, we test experimentally two intensity modulation methods: laser diode gain switching by modulating its pump current, and external intensity modulator connected after the laser diode. The test of pump-current modulation shows a side-channel in the time domain, which allows Eve to distinguish the signal state from the decoy state. We then model a PNS attack that bypasses the decoy state protocol in~\cref{attack}. By optimizing the attack's time windows, Eve should be able to eavesdrop the secret key successfully. On the other hand, from Alice and Bob's point of view, we give a method to analyze the security of the decoy state protocol with imperfect source in~\cref{key rate}. Here the imperfect source means that the signal state and decoy state are partially distinguishable in any degrees of freedom. Two analytic formulas are given to estimate the key rate under such an imperfect source. To improve the secure key rate, an advanced method with calibrated Bob is provided as well. Our method is based on the decoy state protocol with three intensities (a signal state, a weak decoy state, and a vacuum state), which is the most widely used setting. In~\cref{discussion}, our method is applied to three QKD systems to calculate their secure key rate versus distance. We conclude in~\cref{conclusion}.

\section{Decoy state protocol}
\label{basic}

As a fundamental theory of our research, we recap the decoy state protocol first in this section. Here we take the weak~+~vacuum decoy state protocol~\cite{ma2005} as an example to explain the basic idea of the decoy state protocol. This simple one weak~+~vacuum decoy state protocol is commonly used in Bennett-Brassard~1984 (BB84) QKD system~\cite{bennett1984}, as it provides the optimal key rate in the case of only two decoy states~\cite{ma2005}. The security analysis in~\cref{attack,key rate,discussion} also follows this decoy state model.

According to the analysis of Gottesman-Lo-L\"utkenhaus-Preskill (GLLP)~\cite{gottesman2004}, the key rate of QKD with the WCS can be written as
\begin{equation}\label{GLLP}
R \geq q\{-Q_\mu H_2(E_\mu) f(E_\mu) +P_1^\mu Y_1^\mu[1-H_2(e_1^\mu)]\}.
\end{equation}
Here $q=1/2$ for BB84 protocol (if one uses the efficient BB84 protocol~\cite{lo2005a}, $q\approx 1$), the subscript $\mu$ means the intensity of a signal state, $Q_\mu$ ($E_{\mu}$) is the total gain (error rate) of the signal state, $Y_1^\mu$ and $e_1^\mu$ are the yield and error rate of single-photon pulses, $P_1^\mu$ is the probability of single-photon pulses, $f(x)$ is the bidirectional error correction efficiency, normally $f(x)\geq 1$ with Shannon limit $f(x)=1$, and $H_2(x)=-x \log2(x) -(1-x)\log2(1-x)$ is the binary Shannon information entropy.

In~\cref{GLLP}, $Q_\mu$ and $E_\mu$ are directly obtained in an experiment, and $P_1^\mu$ is known for a given source. Thus, the major task of the decoy state is to tightly estimate the  lower bound of $Y_1^\mu$ and upper bound of $e_1^\mu$. Note the fact that, if the phase of the WCS is randomized from 0 to $2\pi$ (the phase randomization assumption), the density matrix of the WCS can be written as
\begin{equation}\label{Rho}
\rho_{\omega}=\sum_{n=0}^{\infty} P_n^\omega |n\rangle\langle n|,
\end{equation}
where $\omega=\{\mu,\nu,0\}$ represents the average intensity of pulse signal state $\mu$, decoy state $\nu$, and vacuum state that is always 0. $P_n^\omega$ is the probability distribution of $n$-photon number from the source with the intensity $\omega$. For the WCS, $P_n^\omega=e^{-\omega}\omega^n/n!$ . Without loss of generality, we assume $\mu>\nu$. Thus, the total gain and error rate can be written as
\begin{equation}\label{QE_calcu}
\begin{split}
Q_\omega&=\sum_{n=0}^\infty P_n^\omega Y_n^\omega,\\
Q_\omega E_\omega&=\sum_{n=0}^\infty P_n^\omega Y_n^\omega e_n^\omega.
\end{split}
\end{equation}
Here $Y_n^\omega$ (or $e_n^\omega$) is the yield (or error rate) given that Alice sends a $n$-photon pulse from the source with intensity $\omega$. Obviously, if Eve does not have any prior information about the intensity of Alice's pulse, we can assume that
\begin{equation}\label{ASSump}
\begin{split}
Y_n^\mu =Y_n^\nu =Y_n,\\
e_n^\mu =e_n^\nu =e_n.
\end{split}
\end{equation}
Then the lower bound of $Y_1$ and upper bound of $e_1$ can be estimated by solving the linear Eqs.~\labelcref{QE_calcu} with weak~+~vacuum decoy states~\cite{ma2005}.

\section{Intensity modulation test}
\label{measurement}

To evaluate the realization of the weak~+~decoy state protocol, we test two intensity modulation methods. The implementation of each has been obtained from a third party, and is tested as supplied, without any tampering or making adjustments.

The first method under testing is the pump-current modulation, similar to Refs.~\onlinecite{yin2017a,liao2017a}. For the signal and weak decoy states, different intensities are produced by applying different pulses of pump current to a laser diode. Thus, the laser diode directly emits optical pulses with different intensities. The vacuum state is generated by turning off the pump current. An optical attenuator then applies a fixed attenuation to all the optical pulses, to reach single-photon level. The second method under testing is an external intensity modulator, similar to Refs.~\onlinecite{rosenberg2007,yuan2007a,dixon2008}. Optical pulses could be produced with a constant intensity from a laser diode first, and then the different intensities of signal and decoy states are modulated by an intensity modulator (IM). Similarly to the former method, a fixed attenuator provides attenuation to the single-photon level.

\begin{figure}
\scalebox{1}{\includegraphics[width=\columnwidth]{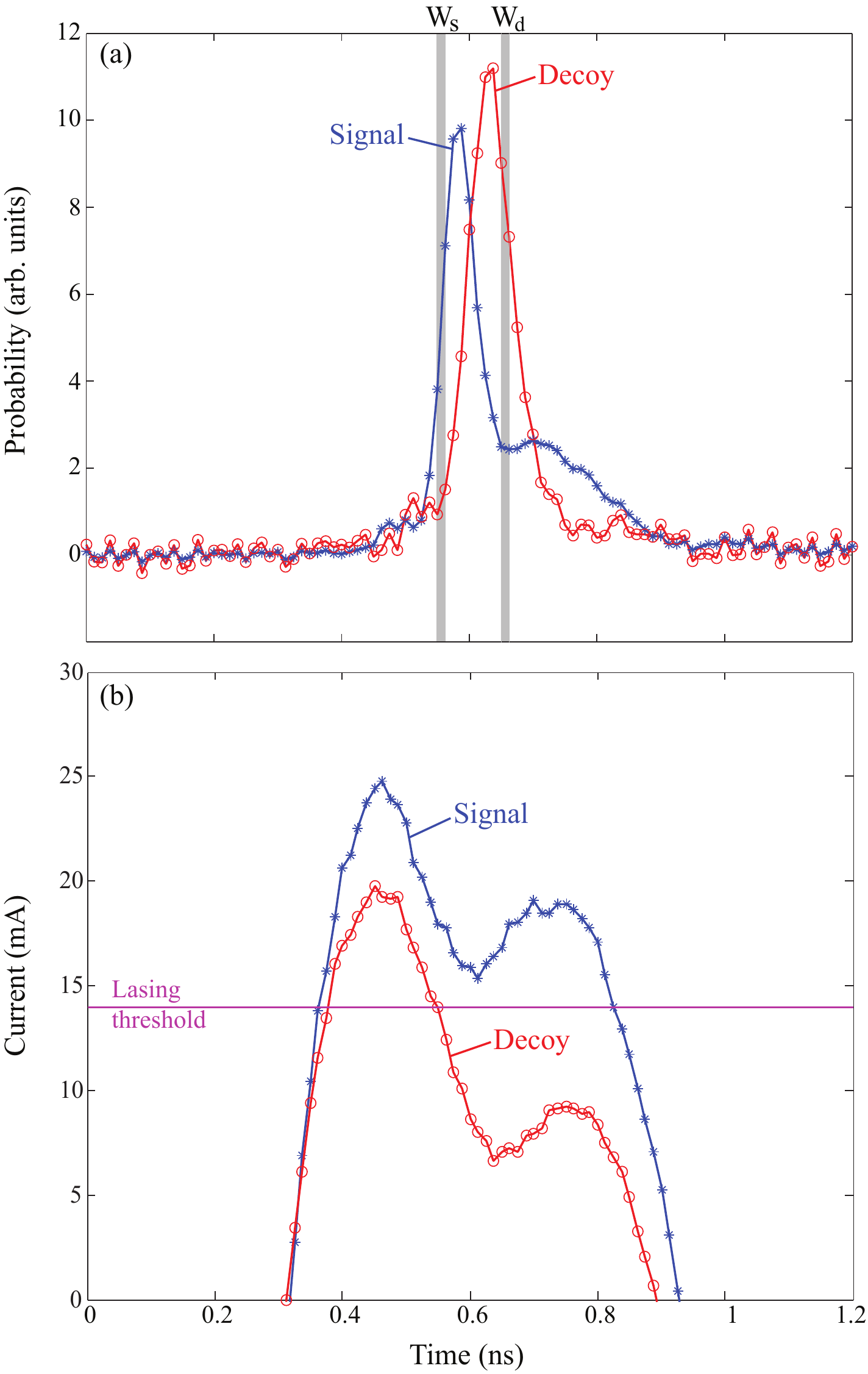}}
\caption{\label{fig:current}(Color online) Pump-current modulation for the laser diode generating $\ket{H}$ polarization state. (a) Normalized intensity distribution of the signal state and the decoy state measured in the time domain. For ease of comparison, the pulses are normalized to have the same area. The original signal-to-decoy intensity ratio is 3:1 ($\mu=0.6, \nu=0.2$). $W_s$ and $W_d$ indicate the typical time windows for Eve to conduct PNS attack, as detailed in~\cref{attack}. (b) Laser-diode's pump current. The current is calculated from the measured voltage across the laser diode module. The lasing threshold, $14~\milli\ampere$, is shown as a line. The relative time alignment between (a) and (b) is a guess.}
\end{figure}

\begin{figure}
\scalebox{1}{\includegraphics[width=\columnwidth]{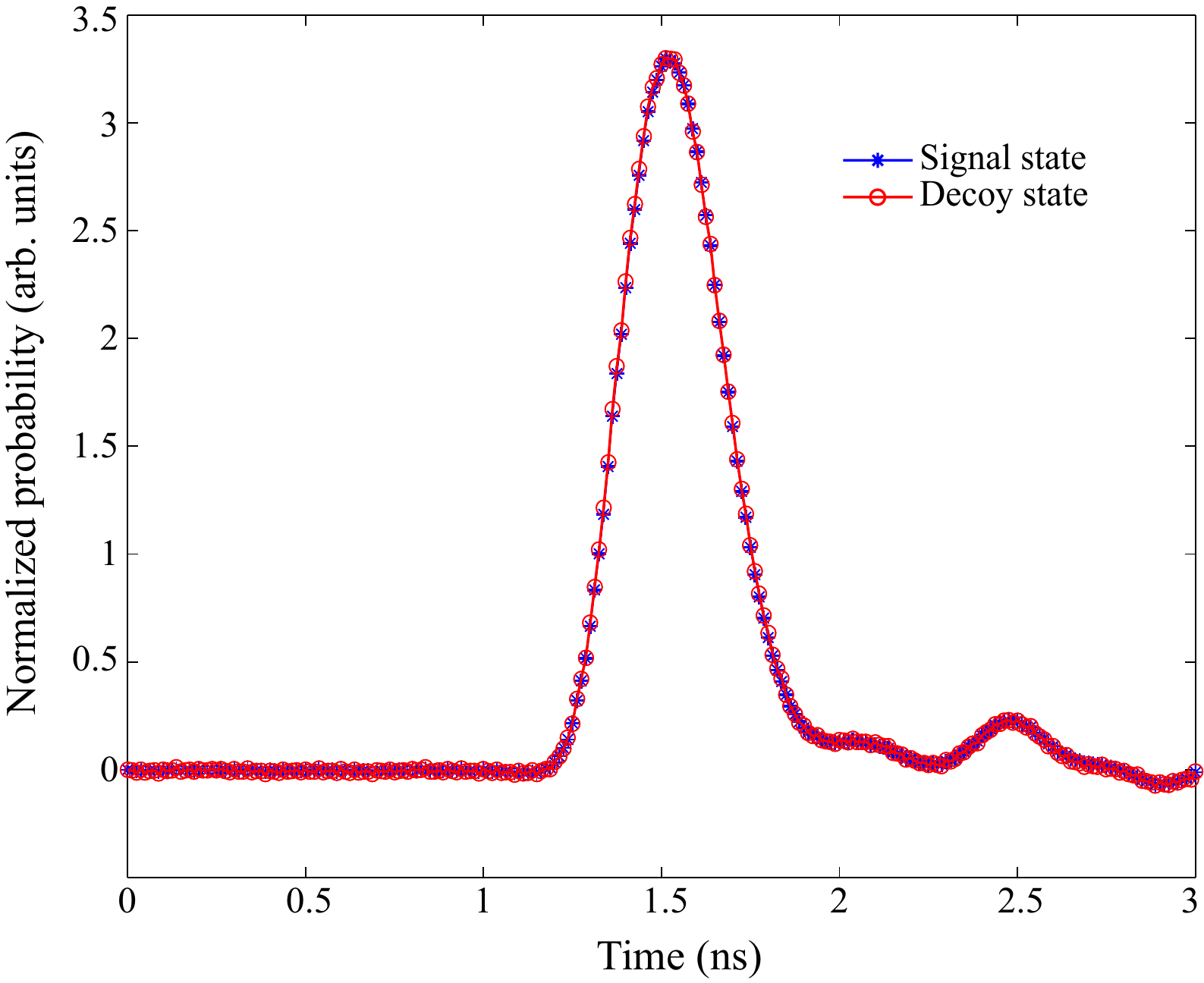}}
\caption{\label{fig:IM}(Color online) External intensity modulation. Normalized intensity distribution of the signal state and the decoy state measured in the time domain. For ease of comparison, the pulses are normalized to have the same area.}
\end{figure}

Our intensity measurement of the optical pulses is taken before the fixed attenuation is applied. The optical pulses are measured by a photodetector ($40~\giga\hertz$ bandwidth) and an oscilloscope ($33~\giga\hertz$ bandwidth), averaging over $\gtrsim 5000$ pulses. We obtain the normalized probability distribution of emitting photons over time which is shown in Figs.~\ref{fig:current}(a) and~\ref{fig:IM}. Although we measure the intensity of classical optical pulses, the probability of emitting single photon should follow the same distribution, because constant attenuation is applied.

For the case of pump-current modulation, we measure the intensities of signal and decoy states for the polarization state $\ket{H}$. \Cref{fig:current}(a) clearly shows that the probability distributions of emitting the signal state and the decoy state do not totally overlap. The main peaks of these two distributions are mismatched. The signal state emits earlier than the decoy state with high probability, and has a secondary peak from $662$ to $937~\pico\second$. Over the same time interval, the probability distribution of the decoy state drops to low values. The timing mismatch of the signal state and the decoy state clearly violates the basic assumption of indistinguishability in the decoy state protocol. As we show numerically in~\cref{attack}, this can be exploited by Eve to bypass the protection of the decoy state protocol. However, the measured result of external intensity modulation in \cref{fig:IM} does not show a measurable timing mismatch between signal and decoy states. This is expected, because the pulse generation and intensity modulation in this type of source are physically decoupled and performed by separate devices. As long as there is no electrical crosstalk between the laser diode driver and intensity modulator driver, no correlation is expected. This is the case, as~\cref{fig:IM} shows.

To investigate the reason for timing mismatch in the case of pump-current modulation, we measure the current flowing through the laser diode (Agilecom WSLS-940010C4123). A differential probe with $30~\giga\hertz$ bandwidth (Agilent N5445A) is used to measure the differential voltage $V$ across the laser diode and its built-in serial resistor $R_s = 20~\ohm$. Since the laser diode forward voltage $V_d = 1.23~\volt$ is known from its test sheet, we can calculate the pump current $I=(V-V_d)/R_s$. This calculated current is shown in~\cref{fig:current}(b).

If the laser diode were pumped by a constant current, any current above the lasing threshold $I_\text{th} = 14~\milli\ampere$ (shown in the figure) would result in continuous-wave (c.w.)\ laser emission. However, when the current is initially zero then rapidly increased above $I_\text{th}$, the diode does not begin to lase instantly~\cite{agrawal1993}. First, a certain number of carriers has to be injected into the p-n junction before the diode reaches population inversion, and that takes time (the higher the current, the less time). Once the population inversion is reached and the diode attains light amplification condition, the few spontaneously emitted photons present in the optical cavity need time to amplify into a strong coherent light. This results in a fraction-of-nanosecond delay between the application of current and the start of strong light emission. In this process, the population inversion and emitted light power briefly overshoot the steady-state. They then undergo a few oscillations with $\sim 100~\pico\second$ period and eventually settle at the steady-state c.w.\ level if the pump current continues~\cite{agrawal1993}. However, if the pump current is interrupted, as is the case with our device under test, the lasing stops. As can be seen in~\cref{fig:current}, the signal state is produced by a higher peak current pulse, the laser begins emitting light earlier and has time to emit two light pulses (i.e.,\ light power oscillations) before the current stops. The decoy state is produced by a lower peak current pulse, light emission begins later and the laser only has time to emit one light pulse. This physics of laser diode operation is well-known to the manufacturers of pulsed laser diodes (e.g.,\ PicoQuant). However, the engineers who selected this modulation method for the QKD system under our test did not initially realise that it created a security loophole.

While the mismatch might be reduced by adjusting the timing and shape of the pump current between the states, it is unlikely to be fully eliminated. The difference in light pulse shapes is due to the physics of laser diode operation. Further testing is needed to find how well the mismatch can be controlled in a practical device.

\section{PNS attack}
\label{attack}

In the case of pump-current modulation, because the signal state and the decoy state are partially distinguishable in the time domain, the PNS attack becomes possible again. Here we consider a special PNS attack summarized in~\cref{tbl:strategy}. Eve selects time windows $W_s$ and $W_d$ to observe states sent by Alice. By properly setting the intervals of $W_s$ and $W_d$, Eve treats all states observed in $W_s$ ($W_d$) as the signal state (the decoy state). Then she performs the PNS attack. For single-photon states, Eve blocks or forwards those that are in the observation windows, while blocks all of those that are out of the observation windows. Once the states contain two or more photons, Eve keeps one photon and either blocks or forwards the rest of the photons to Bob in the observation windows, but forwards all the photons to Bob when the states are out of the windows. If Eve obtains photons in both $W_s$ and $W_d$, she randomly keeps photons in only one window, and forwards the rest of photons to Bob.

By following the criteria of a successful attack proposed in Ref.~\onlinecite{tang2013}, the success ability of the above attack could be analyzed. A successful attack lets Eve know partial information about the final secret key. In other words, Alice and Bob's key remains partially insecure after post-processing. To show this, a lower bound of the key rate under Alice and Bob's estimation, $R^l$, and an upper bound of the key rate under Eve's attack, $R^u$, are compared. If
\begin{equation}\label{cond}
R^l > R^u,
\end{equation}
the shared final key must be partially insecure and Eve knows some amount of information. This is the result Eve would like to reach in her attack.

\renewcommand{\arraystretch}{1.9} 
\begin{table}
  \caption{Hacking strategy and corresponding yields.}
  \label{tbl:strategy}
    \begin{tabular}{c|cc}
       PNS attack &In the time windows & \makecell{Outside the \\ time windows}\\
       \hline
        Single-photon states & Forward or block\footnote{\label{a}Depends on the optimized yield $Z_n^{\omega}$ in simulation.} & Block\\
        Multiphoton states & \makecell{Keep one photon and \\forward or block others\footref{a}} & Forward \\
        Yield $Y_n^{\omega_{\textrm{Eve}}}$ & $Z_n^{\omega}$ & \makecell{0 ($n = 1$) \\ $Y_n^{\omega}$ ($n\geq2$)} \\
       \end{tabular}
\end{table}

The lower bound of the key rate is the one used in the decoy state protocol~\cite{ma2005}:
\begin{equation}\label{lowerbound}
R^l = -Q_{\mu}H_2(E_{\mu}) f(E_\mu) + Y_1^{\mu}\mu e^{-\mu}[1-H_2(e_1^{\mu})],
\end{equation}
which is consistent with~\cref{GLLP} when we consider the efficient BB84 protocol~\cite{lo2005a}, $q = 1$. Here $Y_1^{\mu}$ and $e_1^{\mu}$ are single-photon yield and error rate in the normal decoy state protocol. It is the secure key rate from Alice and Bob's point of view under the attack. Since Alice and Bob do not know about Eve's attack, the estimation of the lower bound of $Y_1^{\mu}$ and the upper bound of $e_1^{\mu}$ still follows the weak~+~vacuum decoy protocol~\cite{ma2005} with the assumption of indistinguishability. The actual upper bound of the key rate under the PNS attack~\cite{tang2013} is
\begin{equation}\label{upperbound}
R^u = Y_1^{\mu_{\textrm{Eve}}}\mu e^{-\mu},
\end{equation}
where $Y_1^{\mu_{\textrm{Eve}}}$ is the real overall yield of single photon states under Eve's attack. The goal of our attack is to minimize the upper bound in~\cref{upperbound} to satisfy inequality~\labelcref{cond}, while matching the value of $Q_{\omega}$ and even reaching lower QBER than $Q_{{\omega}}E_{\omega}$. Then the attack will remain unnoticed.

Based on the measurement result in~\cref{fig:current}(a), Eve can only partially distinguish the signal state and the decoy state. In a certain observation window, we define the following guessing probability. The conditional probability $P(i\big\vert j)$ is defined as Eve guesses the state is $i$ given Alice actually sending the $j$ state. Here $i, j \in [s, d]$, which means $i$ or $j$ is either the signal state, $s$, or the decoy state, $d$. Thus, $P(s\big\vert s)$ and $P(d\big\vert d)$ are the probabilities of correct guess in $W_s$ and $W_d$ respectively, while $P(s\big\vert d)$ and $P(d\big\vert s)$ are the probabilities of wrong guess in the same windows.

As mentioned in the hacking strategy, once Eve observes multiphoton states in $W_s$ or $W_d$, she keeps a single photon and might forward or block the remaining photons to Bob. In order to maintain the statistics of $Q_{\omega}$ and $Q_{{\omega}}E_{\omega}$, Eve has to manipulate detection yield in the observation windows from $Y_n^{\omega}$ to $Z_n^{\omega}$ as shown in~\cref{tbl:strategy}. In the time window $W_s$~($W_d$), the yield is denoted as $Z_n^{\mu}$~($Z_n^{\nu}$). Please note that Eve is allowed to use a lower-loss, or even lossless, channel, which means $Z_n^{\omega}$ could be greater than $Y_n^{\omega}$. At the phase of decoy announcement in QKD protocol, Bob classifies detection slots according to Alice's signal and decoy information. Thus, under Eve's attack, the yields $Y_n^{\omega_{\textrm{Eve}}}$ actually should be recalculated as the following. For the single-photon states, Eve fully controls the yields, since the single-photon states outside the time windows are blocked. Thus, $Y_1^{\omega_{\textrm{Eve}}} $ are given by
\begin{equation}\label{newY1}
\begin{split}
Y_1^{\mu_{\textrm{Eve}}} = P(s\big\vert s)Z_1^{\mu} + P(d\big\vert s)Z_1^{\nu},\\
Y_1^{\nu_{\textrm{Eve}}} = P(s\big\vert d)Z_1^{\mu} + P(d\big\vert d)Z_1^{\nu}. 
\end{split}
\end{equation}
For multiphoton states ($n\geq2$), Eve forwards the states to Bob when these states are outside the observation windows, so $Y_n^{\omega_{\textrm{Eve}}} $ are given by
\begin{equation}\label{newY}
\begin{split}
Y_n^{\mu_{\textrm{Eve}}} = P(s\big\vert s)Z_n^{\mu} + P(d\big\vert s)Z_n^{\nu} + [1 \!-\! P(s\big\vert s) \!-\! P(d\big\vert s)]Y_n^{\mu},\\
Y_n^{\nu_{\textrm{Eve}}} = P(s\big\vert d)Z_n^{\mu} + P(d\big\vert d)Z_n^{\nu} + [1 \!-\! P(s\big\vert d) \!-\! P(d\big\vert d)]Y_n^{\nu}.
\end{split}
\end{equation}
Correspondingly, the overall gains of the signal state and the decoy state are
\begin{equation}\label{newQ}
\begin{split}
Q_{\mu_{\textrm{Eve}}} = Y_0^{\mu_{\textrm{Eve}}}e^{-\mu}  + \sum_{n=1}^{\infty} Y_n^{\mu_{\textrm{Eve}}} e^{-\mu}\frac{\mu^{n}}{n!},\\
Q_{\nu_{\textrm{Eve}}} = Y_0^{\nu_{\textrm{Eve}}}e^{-\nu} + \sum_{n=1}^{\infty} Y_n^{\nu_{\textrm{Eve}}} e^{-\nu}\frac{\nu^{n}}{n!},
\end{split}
\end{equation}
where $Y_0^{\omega_{\textrm{Eve}}}$ are dark count rates under the attack. The overall QBERs are given by
\begin{equation}\label{newQBER}
\begin{split}
E_{\mu_{\textrm{Eve}}}Q_{\mu_{\textrm{Eve}}} = \frac{1}{2}Y_0^{\mu_{\textrm{Eve}}}e^{-\mu} + \sum_{n=1}^{\infty} \frac{1}{2}P(d\big\vert s)Z_n^{\nu} e^{-\mu}\frac{\mu^{n}}{n!},\\
E_{\nu_{\textrm{Eve}}}Q_{\nu_{\textrm{Eve}}} = \frac{1}{2}Y_0^{\nu_{\textrm{Eve}}}e^{-\nu} + \sum_{n=1}^{\infty} \frac{1}{2}P(s\big\vert d)Z_n^{\mu} e^{-\nu}\frac{\nu^{n}}{n!}.
\end{split}
\end{equation}
Here we consider an extreme case. A dark count introduces error half the time. There is no error if signal and decoy states are correctly distinguished by Eve or states are outside the windows $W_s$ and $W_d$. However, wrong guess in $W_s$ and $W_d$ results in random clicks, which introduces error half the time.

According to the standard decoy state protocol~\cite{ma2005}, the normal overall gains should be
\begin{equation}\label{Q}
\begin{split}
Q_{\omega} = Y_0 + 1- e^{-\eta\omega},
\end{split}
\end{equation}
where $Y_0$ is the dark count rate, and $\eta$ is the total transmittance of the QKD system. $\eta$ is given by
\begin{equation}\label{eta}
\eta= \eta_{\textrm{Bob}} 10^{-\alpha L/10},
\end{equation}
where $\eta_{\textrm{Bob}}$ is the transmittance of Bob's optical device, including detector efficiency, and $\alpha$ is the transmittance of channel between Alice and Bob. Typically, $\alpha=0.21~\deci\bel/\kilo\metre$ for the commercial fibre at $1550~\nano\metre$. $L$ is the length of channel. The normal overall QBER should be
\begin{equation}\label{QBER}
\begin{split}
E_{\omega}Q_{\omega} =\frac{1}{2} Y_0 +e_{\textrm{detector}}( 1- e^{-\eta\omega}), 
\end{split}
\end{equation}
where $e_{\textrm{detector}}$ is the probability that a photon goes to erroneous detector, characterizing the alignment and stability of a QKD system.

To achieve a successful attack, the upper bound of the key rate $R^u$ should be minimized, which is equivalent to minimizing $Y_1^{\mu_{\textrm{Eve}}}$ in~\cref{newY1}. Meanwhile, to achieve a traceless attack, the attack has to follow the same detection statistics by optimizing $Z_n^{\mu}$, $Z_n^{\nu}$, $P(s\big\vert s)$, $P(d\big\vert s)$, $P(s\vert d)$ and $P(d\vert d)$ for every distance value. Therefore, it becomes an optimization problem under certain constraints:
\begin{equation}\label{optimization}
\begin{split}
\min_{Z_n^{\mu}, Z_n^{\nu}; P(s\vert s), P(d\vert s)} Y_1^{\mu_{\textrm{Eve}}}
\end{split}
\end{equation}
subject to
\begin{equation}\label{constraint}
\begin{split}
Q_{\mu} = Q_{\mu_{\textrm{Eve}}},\\
Q_{\nu} = Q_{\nu_{\textrm{Eve}}},\\
E_{\mu}Q_{\mu} \geqslant E_{\mu_{\textrm{Eve}}}Q_{\mu_{\textrm{Eve}}},\\
E_{\nu}Q_{\nu} \geqslant E_{\nu_{\textrm{Eve}}}Q_{\nu_{\textrm{Eve}}},\\
Z_n^{\mu}, Z_n^{\nu} \in [0,1],\\
P(s\vert s), P(d\vert s), P(s\vert d), P(d\vert d) \in [0,1].
\end{split}
\end{equation}
Ideally, the detection efficiency could be $100\%$, so the yield $Z_n^{\omega}$ could reach 1. We also remark that the probabilities $P(i\big\vert j)$ are taken from the measured probability distribution of the states sent by Alice in~\cref{fig:current}(a). $P(s\big\vert s)$ and $P(s\big\vert d)$ should be taken from the time window $W_s$; $P(d\big\vert s)$ and $P(d\big\vert d)$ should be taken from the time window $W_d$. Importantly, since every time window could contain several timing intervals, any observation probabilities mentioned above should be a sum over the time window. 

\begin{figure}
\scalebox{1}{\includegraphics[width=\columnwidth]{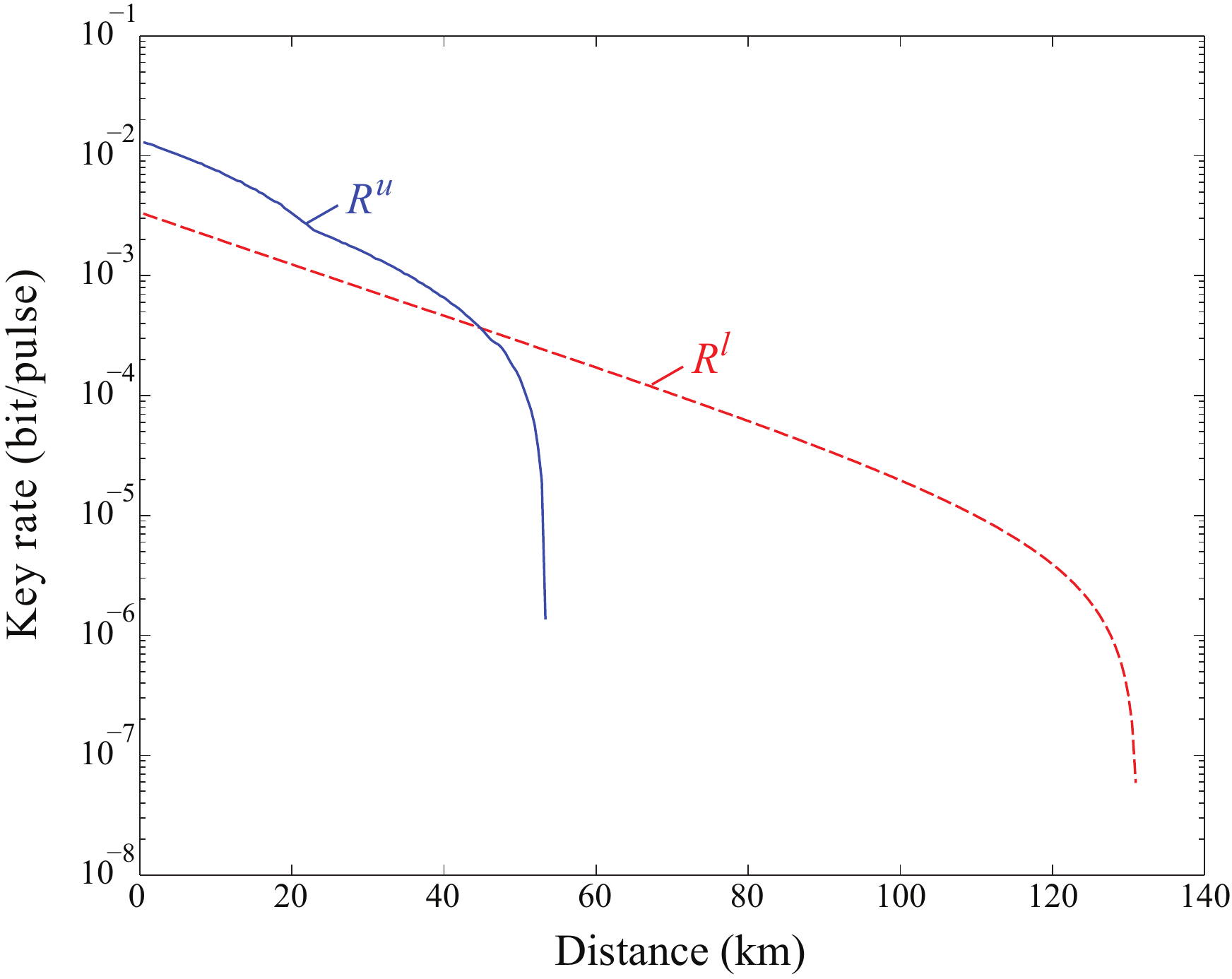}}
\caption{\label{fig:optimization}(Color online) The lower bound $R^l$ and optimized upper bound $R^u$ of key rate under our simulated attack. The detection parameters are taken from GYS experiment~\cite{gobby2004}: the dark count rate $Y_0= 1.7\times 10^{-6}$, the transmission in Bob's apparatus $\eta_{\textrm{Bob}}= 4.5 \%$, the misalignment error rate $e_{\textrm{detector}} = 3.3 \%$ and the error correction efficiency $f(E_\mu) = 1.22$.}
\end{figure}

The simulation result is shown in~\cref{fig:optimization}. To follow the initial analysis of weak~+~vacuum decoy state protocol in Ref.~\onlinecite{ma2005}, we also use the detection parameters from Gobby-Yuan-Shields (GYS) experiment~\cite{gobby2004} in our attack simulation. However, we assume the source has characteristics as in~\cref{fig:current}(a); this source actually comes from a different QKD system with the mean photon number $\mu$ = 0.6 for the signal state and $\nu = 0.2$ for the weak decoy state. According to inequality~\labelcref{cond}, once the optimized upper bound starts becoming smaller than the lower bound, Eve can successfully execute the PNS attack. \Cref{fig:optimization} shows that Eve is able to successfully hack it and eavesdrop some of the secret key when the distance between Alice and Bob is longer than $45~\kilo\metre$. The attack windows $W_s$ and $W_d$ are optimized for every distance point to get the lowest $R^u$ at this distance. For example, when the distance between Alice and Bob is $49~\kilo\metre$, the optimized $W_s$ and $W_d$ are shown as grey zones in~\cref{fig:current}(a).

\section{Secure key rate with imperfect source}
\label{key rate}

The previous section shows the effect of partial distinguishability between signal and decoy states in the time domain. However, the side-channel that partially distinguishes signal and decoy states could be more general. For example, generating signal and decoy states by individual laser diodes is widely employed in QKD systems~\cite{peng2007,yin2008,liu2010}, even in the first quantum satellite~\cite{liao2017}. Unfortunately, it has been shown that this type of state preparation might leak the modulation information in the time and frequency (spectral) domains~\cite{nauerth2009}. For another preparation method of one laser diode with an IM in a plug-and-play system, Eve can shift the arriving time of pulses to the rising edge of intensity modulation, obtaining a side-channel in the frequency domain in the plug-and-play system~\cite{jiang2012}. Moreover, modulation information of IM might be read out by an active Trojan-horse attack~\cite{vakhitov2001,tamaki2016}. Even if the intensity modulation is perfect, the laser pulses with non-random phases give Eve a chance to distinguish signal and decoy states~\cite{tang2013}. Therefore, it is important to build a general security model that tolerates such side-channels. In this section, we modify the model of the decoy state protocol to consider such imperfect sources, and derive two analytic formulas that estimate the contribution of single-photon pulses.

\subsection{Model}
\label{sec:keyrate-model}

We analyze the weak~+~vacuum decoy state protocol with intensities $\omega=\{\mu,\nu,\nu_1\}$. Without loss of generality, we assume $\mu>\nu>\nu_1$ and $\mu>\nu+\nu_1$. Here $\nu_1$ represents the intensity of a vacuum state. 

If the imperfection of source is taken into account, the density matrix of Alice's states [\cref{Rho}] can be rewritten as
\begin{equation}\label{Rho_2}
\rho'_{\omega}=\rho_\omega\otimes \rho_\omega(\lambda)=\sum_{n=0}^{\infty} \int_\lambda d\lambda P_n^\omega  f_\omega (\lambda) |n,\lambda\rangle\langle n,\lambda| .
\end{equation}
Here $\rho_\omega(\lambda)$ is the quantum state used by Eve to distinguish the signal state and the decoy state for each pulse. Note that $\rho_\omega(\lambda)$ can be an actual quantum state, or any additional dimension of Alice's pulses. $\lambda$ represents the parameter (for example, the time, frequency and so on) measured by Eve, which is used to distinguish the signal state or the decoy state. $f_\omega(\lambda)$ is the normalized probability distribution of $\lambda$ ($\int_\lambda f_\omega(\lambda) d\lambda=1$), which depends on the intensities of Alice's pulse $\omega$. Note that if $\lambda$ is a discrete variable, $\int d\lambda$ should be changed to $\sum_\lambda$ in~\cref{Rho_2}. We assume that the measured parameter of Eve, $\lambda$, is independent of the photon number of Alice's state, $n$. I.e., $|n,\lambda\rangle=|n\rangle|\lambda\rangle$. Obviously, if $\rho_\omega(\lambda)$ is independent on the intensities of Alice's pulse, which means $\rho_\mu(\lambda)=\rho_\nu(\lambda)=\rho_{\nu_1}(\lambda)\equiv \rho(\lambda)$, then \cref{Rho_2} becomes~\cref{Rho}. Thus, the general decoy state method can be used to estimate the bound of yield and error rate for the single photon pulses~\cite{ma2005}.

By combining Eqs.~\labelcref{QE_calcu} with~\cref{Rho_2}, the total gain and error rate of Alice's states should be rewritten as
\begin{equation}\label{QE_calcu2}
\begin{split}
Q_\omega&=\sum_{n=0}^\infty P_n^\omega Y_n^\omega =\sum_{n=0}^\infty P_n^\omega \sum_{\lambda} f_{\omega}(\lambda)Y_n(\lambda),\\
Q_\omega E_\omega&=\sum_{n=0}^\infty P_n^\omega Y_n^\omega e_n^\omega =\sum_{n=0}^\infty P_n^\omega \sum_{\lambda} f_{\omega}(\lambda)Y_n(\lambda)e_n(\lambda),
\end{split}
\end{equation}
where $Y_n(\lambda)$ and $e_n(\lambda)$ are the yield and error rate given that Alice sends an $n$-photon pulse and Eve obtains $\lambda$ in her measurement. Thus $Y_n(\lambda)$ and $e_n(\lambda)$ depend on the parameter $\lambda$, but are independent on the intensities of Alice's pulses $\omega$.

As Ref.~\onlinecite{tamaki2016} mentioned, the imperfection of source can be characterized by the distance between $\rho_\omega(\lambda)$ and $\rho_{\omega'}(\lambda)$, which is given by
\begin{equation}\label{Duv_rho}
D_{\omega\omega'}=\frac{1}{2}tr|\rho_\omega-\rho_{\omega'}|.
\end{equation}
Here $tr{|x|}$ is the trace distance of quantum state.

\bigskip 

\subsection{Lower bound of $\bm{Y_1^\mu}$}
\label{sec:keyrate-lowerbound}

Now we derive the lower bound of $Y_1^\mu$ according to the model given above. From~\cref{Duv_rho}, it is easy to obtain inequalities
\begin{equation}\label{Yen_bound}
\begin{split}
|Y_n^{\omega}-Y_n^{\omega'}|\leq 2D_{\omega \omega'},\\
|Y_n^{\omega}e_n^{\omega}-Y_n^{\omega'}e_n^{\omega'}|\leq 2D_{\omega\omega'},
\end{split}
\end{equation}
where $\omega, \omega' = \mu, \nu, \nu_1$ and $1>\mu>\nu>\nu_1>0$. Note that by following the proof of Ref.~\onlinecite{tamaki2016}, the factor 2 in~\cref{Yen_bound} may be removed to improve the key rate.

To estimate the lower bound of $Y_1^\mu$, we follow the procedure in Ref.~\onlinecite{ma2005}. We assume that $Y_0^{\mu} =Y_0^{\nu}=Y_0^{\nu_1}=Y_0$, since there is no difference for the vacuum pulse and Eve cannot get any information from such pulse. The lower bound of $Y_1^\mu$ is given by (see \cref{A_Y1} for derivation)
\begin{equation}\label{Y12}
\begin{split}
Y_1^\mu \geq\ &\frac{\mu[e^{\nu}Q_{\nu}\!-\!e^{\nu_1}Q_{\nu_1}\!-\!\frac{\nu^2\!-\!\nu_1^2}{\mu^2}(e^{\mu}Q_{\mu\!}-\!Y_0^L)]}{\mu(\nu-\nu_1)-(\nu^2-\nu_1^2)}  -g(\mu,\nu,\nu_1)\\
\equiv\ &G(\mu,\nu,\nu_1) -  g(\mu,\nu,\nu_1).
\end{split}
\end{equation}
Here,
\begin{equation}
g(\mu,\nu,\nu_1)\equiv \frac{2\mu[D_{\mu\nu}  (e^{\nu}-1) + D_{\mu\nu_1}(e^{\nu_1}-1)]}{\mu(\nu-\nu_1)-(\nu^2-\nu_1^2)}.
\end{equation}
It is easy to check that $G(\mu,\nu,\nu_1)$ is the same as the lower bound of $Y_1^\mu$ in the standard decoy-state protocol (Eq.~(21) in Ref.~\onlinecite{ma2005}), which represents the yield of the single photon pulse with a perfect source. Thus, $g(\mu,\nu,\nu_1)$ represents the leaked information due to the imperfection of source. 

In general decoy state experiments, weak~+~vacuum state protocol is used, which means $\nu_1=0$. Then~\cref{Y12} can be written as
\begin{equation}\label{Y1}
Y_1^\mu \! \geq\! \frac{\mu}{\mu\nu\!-\!\nu^2} [e^{\nu}Q_{\nu}\!-\!\frac{\nu^2}{\mu^2}e^{\mu}Q_{\mu}\!-\!\frac{\mu^2\!-\!\nu^2}{\mu^2}Q_\text{vac}\!-\!2 D_{\mu\nu}(e^{\nu}\!-\!1)].
\end{equation}

\subsection{Upper bound of $e_1^\mu$}
\label{sec:keyrate-upperbound}
The upper bound of $e_1^\mu$ can be estimated by (see \cref{A_e1} for derivation)
\begin{equation}\label{e12}
e_1^\mu \leq \min\{K^\mu, K^\nu, K^{\nu_1}, K^{\mu\nu}, K^{\mu\nu_1}\},
\end{equation}
where
\begin{equation}
\begin{split}
K^\mu = \ & \frac{e^{\mu}Q_\mu E_\mu-e_0 Y_0^L}{\mu Y_1^\mu},\\
K^\nu = \ & \frac{e^\nu Q_\nu E_\nu -e_0 Y_0^L +2\nu D_{\mu\nu}}{\nu Y_1^\mu},\\
K^{\nu_1} = \ & \frac{e^{\nu_1} Q_{\nu_1} E_{\nu_1} -e_0 Y_0^L +2\nu_1 D_{\mu\nu_1}}{\nu_1 Y_1^\mu},\\
K^{\mu\nu} = \ & \frac{e^\mu Q_\mu E_\mu -e^\nu Q_\nu E_\nu +2 D_{\mu \nu}(e^\nu-1)}{(\mu-\nu)Y_1^\mu},\\
K^{\mu\nu_1} = \ & \frac{e^\mu Q_\mu E_\mu -e^{\nu_1} Q_{\nu_1} E_{\nu_1} +2 D_{\mu \nu_1}(e^{\nu_1} -1)}{(\mu-\nu_1)Y_1^\mu}. \nonumber
\end{split}
\end{equation}

When $D_{\mu \nu_1} = 0$, \cref{e12} becomes as the same as the upper bound of $e_1^\nu$ in the standard decoy-state protocol (Eq.~(25) in Ref.~\onlinecite{ma2005}).

For weak~+~vacuum decoy state protocol, $\nu_1=0$ and $Y_0 = Q_\text{vac}$. Then, the upper bound of $e_1^\mu$ can be rewritten as
\begin{equation}\label{e1}
e_1^\mu \leq \min\{K^\mu, K^\nu, K^{\mu\nu}\}. 
\end{equation}

\subsection{Numerical simulation}
\label{sec:keyrate-numerics}

\begin{figure}
\scalebox{1}{\includegraphics[width=\columnwidth]{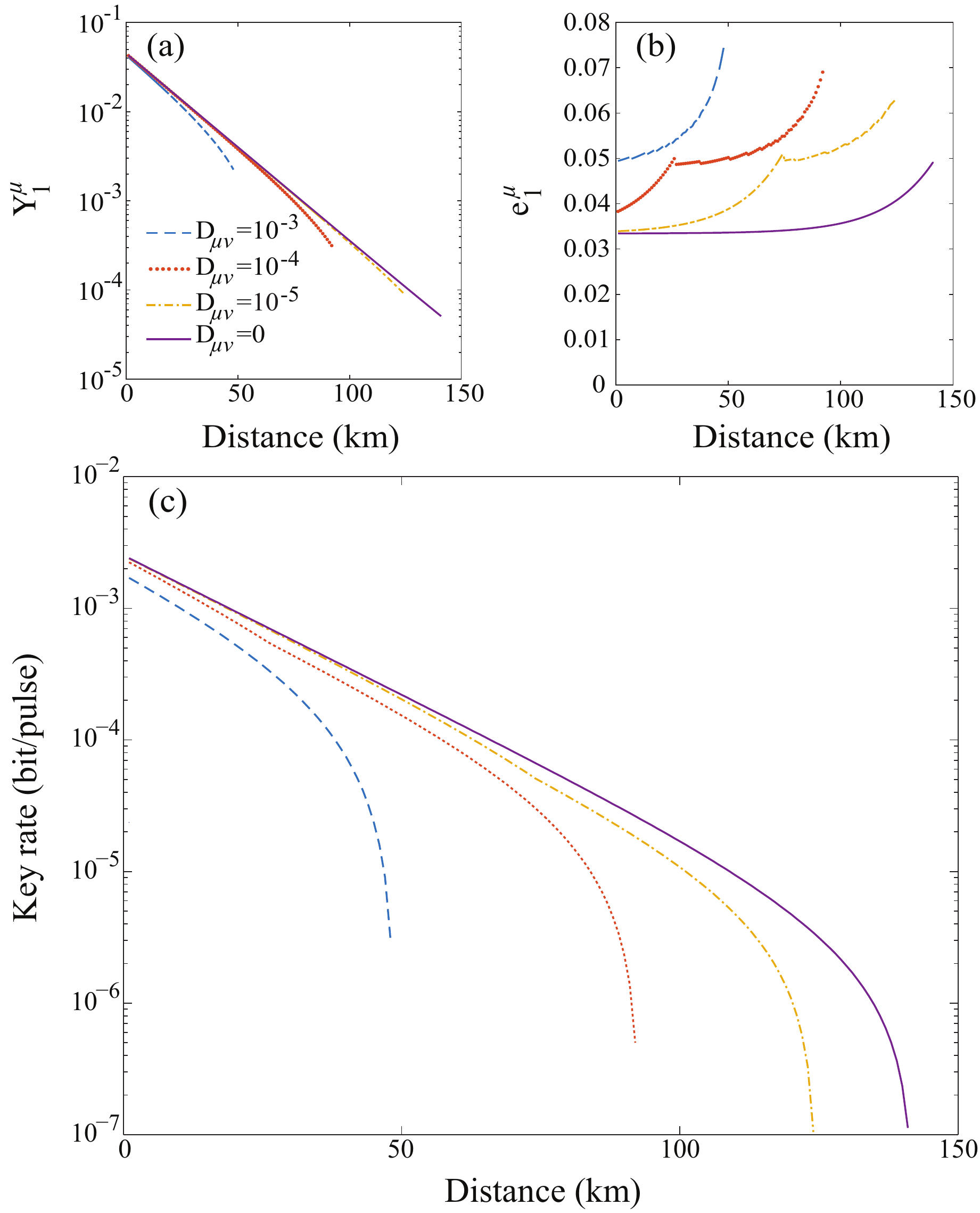}}
\caption{\label{fig:keyrate}(Color online) Estimated system parameters with imperfect source. (a) yield and (b) error rate of the signal state for the single photon pulse, and (c) key rate are shown for different amounts of imperfection $D_{\mu\nu}$. The detection parameters used in the simulations are the same as those in~\cref{fig:optimization}. The intensities of the signal state and the decoy state are optimized with step 0.01 from $\mu \in [0.01,0.5]$, $\nu\in [0.01,0.2]$. We only show the estimated $Y_1^\mu$ and $e_1^\mu$ where the final key rate is positive. No secure key can be generated for $D_{\mu\nu}=10^{-1}$ and $10^{-2}$.}
\end{figure}

We simulate weak~+~vacuum decoy state protocol. When Eve is absent, the total gain and error rate of the signal state and the decoy state are given by~\cref{Q,QBER}. The rest of parameters follow the definition in standard decoy state protocol~\cite{ma2005}. Submitting~\cref{Q,QBER} into~\cref{Y1,e1}, we can obtain the lower bound of yield and the upper bound of error rate for the single photon pulse as shown in~\cref{fig:keyrate}(a) and (b). Then the estimated key rate is given in~\cref{fig:keyrate}(c). This clearly shows that the imperfection of source will reduce the key rate between Alice and Bob rapidly. For example, when the source is perfect ($D_{\mu\nu} = 0$), the maximum distance is about $141~\kilo\metre$. The secure key rate estimated by our method matches that estimated by the original decoy-state protocol in Ref.~\onlinecite{ma2005}, which confirms the estimation is tight. However, the maximum distances are reduced to $124, 92,48~\kilo\metre$ for $D_{\mu\nu}=10^{-5}, 10^{-4}, 10^{-3}$. No positive key rate is possible at any distance for $D_{\mu\nu}=10^{-2}, 10^{-1}$.

\subsection{Theory improvement}
\label{sec:keyrate-improved}

From the modified security proof in the last section, even for the relatively small imperfection $D_{\mu\nu}=10^{-3}$, the maximum secure distance drops quickly from $141~\kilo\metre$ to $48~\kilo\metre$. Here, we propose an advanced security proof to improve the final key rate with the imperfect source by setting a reasonable assumption. Then we could loosen the security constraint when estimating $Y_1^\mu$ and $e_1^\mu$, theoretically improving the secure key rate and the maximum secure distance. 

\begin{figure}
\scalebox{1}{\includegraphics[width=\columnwidth]{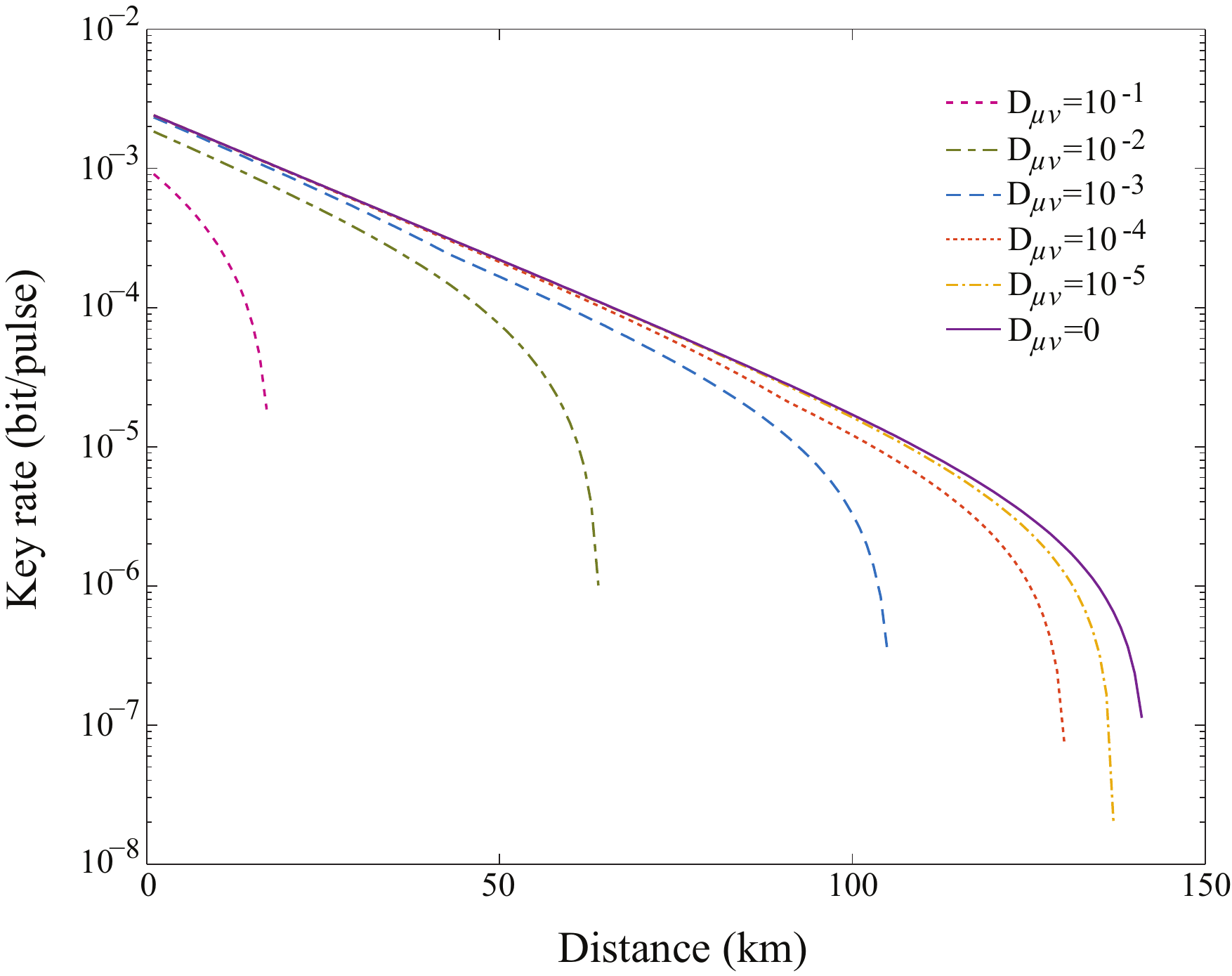}}
\caption{\label{fig:keyrate2}(Color online) Estimated key rate assuming calibrated transmittance in Bob's optical devices. The detection parameters used here are the same as those in~\cref{fig:optimization}.}
\end{figure}

In practical QKD systems based on prepare-and-measure protocol, Bob's devices are located within his protected zone. Thus, it is possible for Bob to calibrate the optical transmittance of his optical devices. Note that here we do not mean that Eve could not change the parameters of Bob's system (for example, change the SPD from Geiger mode to linear mode by performing the blinding attack), but mean that Bob could actively calibrate the transmittance of his partial or all devices. In fact, this assumption has been used to secure the single photon detector of Bob~\cite{maroy2010,maroy2017}. Thus, we think this assumption is reasonable and practical. We can then obtain
\begin{equation}
Y_n(\lambda) \leq 1-(1-\eta_{\textrm{Bob}}^{\textrm{cal}})^n.
\end{equation}
Please note that $\eta_{\textrm{Bob}}^{\textrm{cal}}$ is the calibrated transmittance in Bob, which should be equal with or lower than the total transmittance of Bob $\eta_{\textrm{Bob}}$. In the simulation, we could assume that Bob can calibrate the whole transmittance in his system, thus we could have $\eta_{\textrm{Bob}}^{\textrm{cal}} =\eta_{\textrm{Bob}}$. Then \cref{Yen_bound} can be rewritten as
\begin{equation}\label{Yen_bound2}
\begin{split}
|Y_n^{\omega}-Y_n^{\omega'}|\leq 2D_{\omega \omega'}[1-(1-\eta_{\textrm{Bob}})^n],\\
|Y_n^{\omega}e_n^{\omega}-Y_n^{\omega'}e_n^{\omega'}|\leq 2D_{\omega\omega'}[1-(1-\eta_{\textrm{Bob}})^n].
\end{split}
\end{equation}
Then in weak~+~vacuum decoy state protocol, it is easy to check that the lower bound of $Y_1^\mu$ [\cref{Y1}] and the upper bound of $e_1^\mu$ [\cref{e1}] can be rewritten as
\begin{equation}
\label{modified}
\begin{split}
Y_1^\mu \geq\ & \frac{\mu}{\mu\nu-\nu^2} [e^{\nu}Q_{\nu}-\frac{\nu^2}{\mu^2}e^{\mu}Q_{\mu}-\frac{\mu^2-\nu^2}{\mu^2}Y_0\\
&-2 D_{\mu\nu}(e^{\nu}-e^{\nu(1-\eta_{\textrm{Bob}})})],\\
e_1^\mu \leq\ &\min\{K^\mu,K^\nu,K^{\mu \nu}\},\textrm{where}\\
K^\mu =\ &\frac{e^{\mu}Q_\mu E_\mu-e_0 Y_0}{\mu Y_1^\mu},\\
K^\nu =\ &\frac{e^\nu Q_\nu E_\nu -e_0 Y_0 +2\nu D_{\mu\nu}\eta_{\textrm{Bob}}}{\nu Y_1^\mu},\\
K^{\mu \nu} =\ &\frac{e^\mu Q_\mu E_\mu -e^\nu Q_\nu E_\nu +2 D_{\mu \nu}(e^\nu-e^{\nu(1-\eta_{\textrm{Bob}})})}{(\mu-\nu)Y_1^\mu}.
\end{split}
\end{equation}

Then we could estimate the final key rate with the same method given above. The estimation result in~\cref{fig:keyrate2} clearly shows that when the transmittance of Bob's optical devices ($\eta_{\textrm{Bob}}=4.5 \%$) is taken into account, the final key rate and the maximum distance are improved. For example, in the case that $D_{\mu\nu}=10^{-3}$, the maximum distance increases to $105~\kilo\metre$ from $48~\kilo\metre$. Also note that for $D_{\mu\nu}=10^{-2}$ and $10^{-1}$, the improved proof provides positive key rate up to $64$ and $18~\kilo\metre$. We remark that the assumption of calibrated transmission loss for Bob's devices is not applicable to measurement-device-independent QKD~(MDI QKD), in which the detection part is not in the protected zone and can be fully controlled by Eve.

\section{Discussion and application examples}
\label{discussion}

In~\cref{key rate}, the method used to estimate $Y_1^\mu$ and $e_1^\mu$ in a decoy state protocol considers a type of imperfect source which could partially distinguish signal and decoy states in any degrees of freedom. Once these imperfections are experimentally measured, this method could provide a standard way to calculate the final key rate under such imperfections. Note that this method focuses on the imperfect modulation of signal and decoy states, but does not handle the distinguishability among different BB84 states. We currently assume the identical mismatch of signal and decoy states for each BB84 state. Removing this theoretical limitation could be future work.

Another limitation lies in our experiment. We have measured the distinguishability between signal and decoy states only in the time domain. However, the two modulation methods we have tested might also introduce time-dependent spectral mismatch, which we have not measured. For the gain-switched semiconductor laser, a short pulse usually has a so-called chirp, a fast-changing wavelength modulation~\cite{koch1984,linke1985}. The spectral and intensity modulation contribute simultaneously to the distinguishability, resulting in a joint distribution of $D_{\mu\nu}$ as explained later in this section. The external intensity modulator may also affect the spectrum of pulses~\cite{koyama1988,kawanishi2001}. However, the requisite time-resolved spectroscopy is a more complex measurement~\cite{linke1985,saunders1994,niemi2002}, which could be investigated in the future. For the two devices tested, we henceforth assume distinguishability in the time domain only. 

\begin{figure}
\scalebox{1}{\includegraphics[width=\columnwidth]{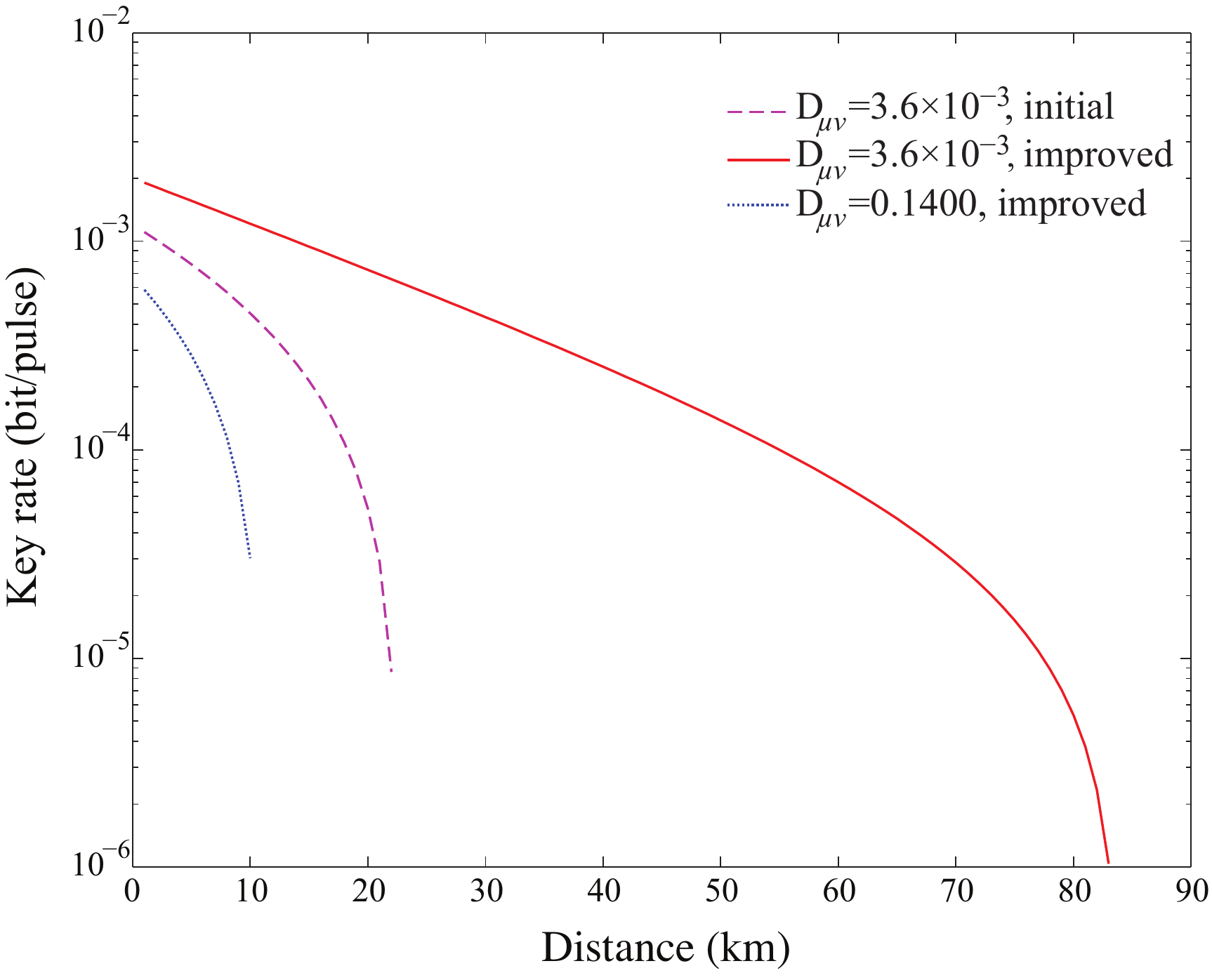}}
\caption{\label{fig:real key} (Color online) Estimated key rate for different experimental distinguishability of signal and decoy states. The detection parameters used here are the same as in~\cref{fig:optimization}. The key rate is estimated by the initial proof with~\cref{Y1,e1} and also the improved proof with~\cref{modified}. Note that in the case of $D_{\mu\nu}=0.1400$, no secure key can be generated with the initial proof, but the positive key rate is possible with our improved proof. $D_{\mu\nu}=0.4005$ cannot generate positive key rate in either proof.}
\end{figure}

We now apply our security proof to the measurement results of the two sources tested in~\cref{measurement}, and to one more published source measurement in Ref.~\onlinecite{nauerth2009}. Both the initial proof in~\cref{sec:keyrate-numerics} and the advanced proof in~\cref{sec:keyrate-improved} are applied in each case. The purpose of this application is quantifying the imperfection of signal and decoy states preparation, and showing its effect on the secure key rate. In order to compare the three source implementations, we arbitrarily assume that these different sources are used in the same fibre-based QKD system with GYS parameters at the detection side. The resulting secure key rates are shown in~\cref{fig:real key}.

For the first case shown in~\cref{fig:current}(a), the corresponding value of $D_{\mu\nu}$ given by~\cref{Duv_rho} is 0.4005. Then both our security proofs are applied. For the improved proof, we assume $\eta_{\textrm{Bob}}=4.5 \%$. The secure key rate is \textit{zero} under either key rate estimation. This shows that the modulation imperfection can make the system insecure.

On the contrary, the value of $D_{\mu\nu}$ for the case in~\cref{fig:IM} is only $3.6\times10^{-3}$. This non-zero value probably stems from the noise in our characterization apparatus. Nevertheless, we have to conservatively treat all mismatch as belonging to the source under test. This non-zero value of $D_{\mu\nu}$ still indicates a certain degree of mismatch. As shown in~\cref{fig:real key}, under the initial proof, the maximum distance drops to $22~\kilo\metre$, while under the advanced proof with $\eta_{\textrm{Bob}}=4.5 \%$, it improves to $83~\kilo\metre$. I.e.,\ owing to the much lower mismatch in this case, the positive key rate could be generated. However, the maximum transmission distance is sensitive even to such low mismatch values.

\begin{figure}
\scalebox{1}{\includegraphics[width=\columnwidth]{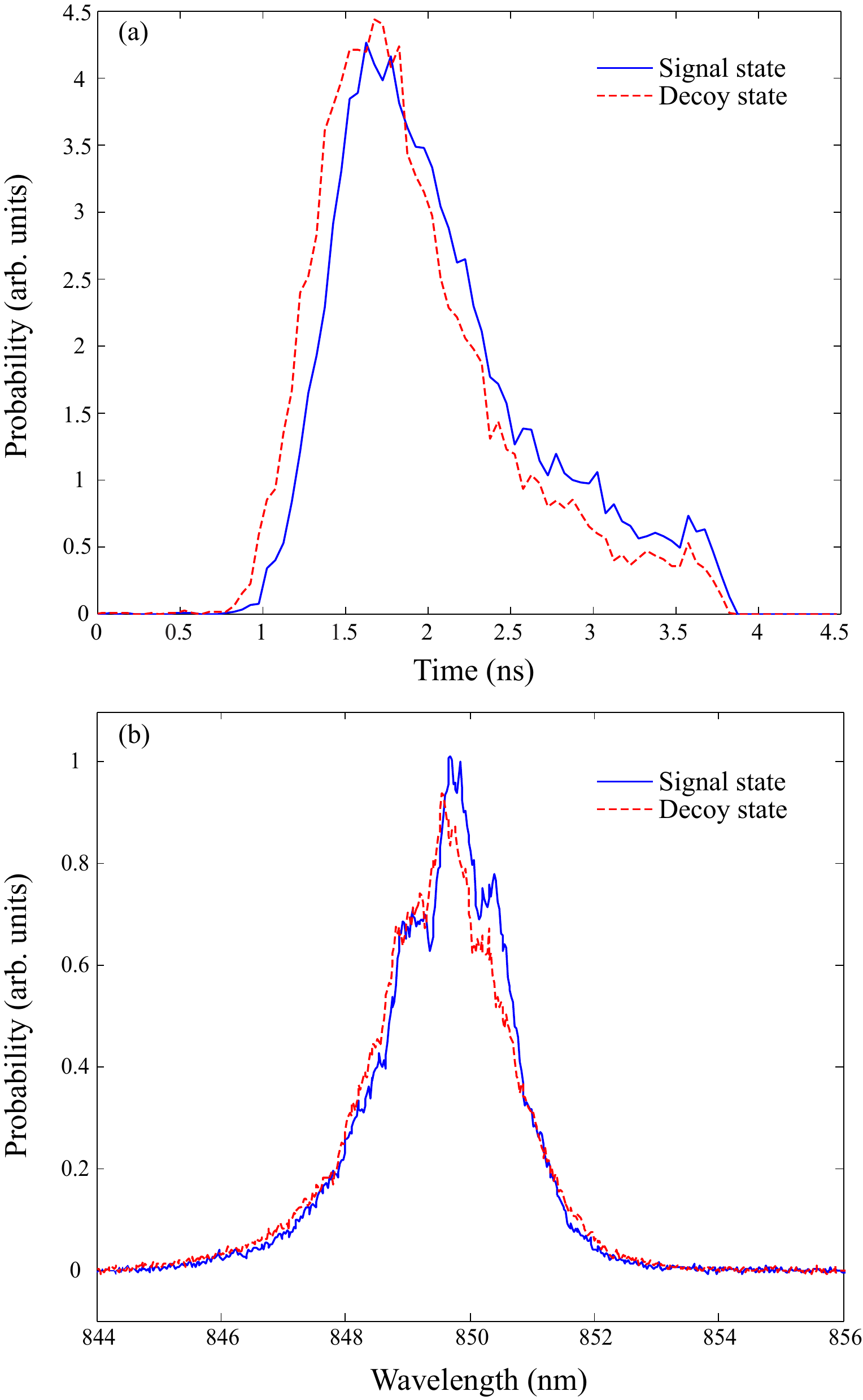}}
\caption{\label{fig:reprint} (Color online) Initial mismatch of signal and decoy states for the vertical-polarization pair of laser diodes in (a) time domain and (b) frequency domain. The timing mismatch can be reduced by tuning electrical delay lines in the laser diodes' driver. Data reprinted from Ref.~\onlinecite{nauerth2009}.}
\end{figure}

Another case of imperfect preparation for signal and decoy states is published in Ref.~\onlinecite{nauerth2009}. In that study, the signal and decoy states are generated by individual laser diodes, which is a common technique~\cite{peng2007,yin2008,liu2010,liao2017}. It shows that mismatches between signal and decoy states are both in the time domain and frequency domain for each individual BB84 state~\cite{nauerth2009}. Because our proof cannot handle the BB84 states individually, we have chosen a typical mismatch between the signal and decoy states in vertical polarization as reprinted in~\cref{fig:reprint}, and assumed arbitrarily that the other three BB84 polarization states have the mismatch identical to that. Even though Ref.~\onlinecite{nauerth2009} studies an imperfect source in a free-space QKD system, we remark that it is reasonable to expect mismatch for any QKD implementations that generate signal and decoy states by individual laser diodes~\cite{peng2007,yin2008,liu2010,liao2017}. Please note that~\cref{fig:reprint} illustrates the initial mismatches between two independent lasers. The timing mismatch can be reduced by adjusting the delay between the laser diodes~\cite{nauerth2009,rau2015}.

The security proof in~\cref{key rate} is able to handle mismatch in arbitrary degrees of freedom, because we do not specify the dimensions of the probability $f_\omega(\lambda)$. $f_\omega(\lambda)$ can be a joint probability. For example, the joint probability distribution of $\omega$ state in the time and frequency domains can be $f_\omega(t, f)$, where $t$ represents the time domain and $f$ represents the frequency domain. Thus, $D_{\mu\nu}$ can be defined as
\begin{equation}\label{Duv_mul}
D_{\mu\nu}=\frac{1}{2}\sum_t \sum_f |f_\mu(t,f)-f_{\nu}(t,f)|.
\end{equation}
Similarly, the calculation of $D_{\mu\nu}$ can be expanded to more than two dimensions. In the specific case shown in~\cref{fig:reprint}, time-resolved spectroscopy necessary to measure the joint probability was not performed. It has been arbitrarily assumed instead that the probability distributions in the time and frequency domains are independent, with a remark that this will need to be verified experimentally~\cite{nauerth2009}. Then, $f_\omega(t, f) = f_\omega(t) f_\omega(f)$ can be calculated from the available experimental data. The corresponding $D_{\mu\nu}$ is 0.1400. With such value of $D_{\mu\nu}$, the initial security proof cannot generate positive key rate for any distance, while the improved proof with $\eta_{\textrm{Bob}}=4.5 \%$ could generate the secure key up to only $10~\kilo\metre$, as shown in~\cref{fig:real key}. 

According to the above analysis and comparison of the three cases, the external intensity modulator shows the smallest mismatch between signal and decoy states, resulting in the highest key rate and longest transmission distance. However, Trojan-horse attack could read out the modulation information from the intensity modulator~\cite{vakhitov2001,tamaki2016}. Thus, countermeasures against Trojan-horse attack are also necessary~\cite{lucamarini2015,sajeed2015}. The use of intensity modulator may also result in non-zero vacuum state, which can be handled by our security proof. The secure key rate can then be estimated by applying~\cref{Y12,e12}.

All the measurement results shown above contain a measurement error. The error may come from the thermal noise of electronic devices, the nonlinearity of optical-to-electrical converters and digital-to-analog converters in the oscilloscope. We have simply treated the measured $D_{\mu\nu}$ as the real mismatch. Thus, the key rates shown in~\cref{fig:real key} are conservative estimates. It is an open question how to extract the real parameter from the noisy test results. 

\section{Conclusion}
\label{conclusion}

In this Article, we have investigated the imperfect sources in QKD systems that implement the decoy state protocol. By testing two intensity modulation methods, we have found that the basic assumption about indistinguishability of signal and decoy states does not hold in practice, especially in the case of laser diode pump current modulation. This pump-current modulation shows timing mismatch between the signal and decoy states. We have modeled a PNS attack based on the timing mismatch that breaks the security of the QKD system. To make the system robust against this loophole, we have considered the distinguishability of signal and decoy states in the security proof, and obtained two analytical formulas to estimate the yield and error rate of the single photon pulses. The result shows that the distinguishability would reduce the secure key rate. Fortunately, the key rate under such imperfection can be improved by calibrating the transmittance of Bob's unit. We have applied this method to three implementations of the decoy state protocol to estimate their secure key rate, which in some cases has become reduced (also limiting the transmission distance), and in some just zero.

The estimation of $Y_1^\mu$ and $e_1^\mu$ with the distinguishable decoy state provides a method to guarantee the security of practical quantum cryptography systems. This method could be employed as a standard tool to estimate the secure key rate, once the distinguishability in the decoy state protocol is quantified in all degrees of freedom. A conceptually similar evaluation method has been proposed for the Trojan-horse attack~\cite{lucamarini2015}. Please note that the key rate shown in this work is based on prepare-and-measure QKD. For MDI QKD and the other quantum cryptography systems, the secure key rate and the other security properties under such imperfect source model should be derived separately.

\acknowledgments
We thank Y.~Zhou for helpful discussions, S.~Nauerth and H.~Weinfurter for providing the raw data from Ref.~\onlinecite{nauerth2009}. This work was funded by the National Natural Science Foundation of China (grant numbers 11674397 and 61601476) and NSERC of Canada (programs Discovery and CryptoWorks21).

\appendix

\section{Estimating the lower bound of $\bm{Y_1^\mu}$}
\label{A_Y1}
Firstly, the lower bound of background rate $Y_0$ is estimated by the following.
\begin{widetext}
\begin{equation}\label{Y0_bound1}
\begin{split}
&\nu e^{\nu_1}Q_{\nu_1}-\nu_1 e^{\nu}Q_{\nu}=\sum_{n=0}^{\infty} \frac{1}{n!}(\nu \nu_1^nY_n^{\nu_1} -\nu_1 \nu^n Y_n^{\nu})\\
=\ & (\nu-\nu_1)Y_0 +  \sum_{n=1}^{\infty} \frac{\nu \nu_1}{n!}(\nu_1^{n-1}Y_n^{\nu_1} -\nu^{n-1} Y_n^{\nu})\\
\leq\ & (\nu-\nu_1)Y_0 + \sum_{n=1}^{\infty} \frac{\nu \nu_1}{n!}[\nu_1^{n-1}(Y_n^{\mu} +2D_{\mu \nu_1}) -\nu^{n-1} (Y_n^{\mu}-2D_{\mu \nu})]\\
=\ & (\nu-\nu_1)Y_0 + \sum_{n=1}^{\infty} [2D_{\mu \nu_1} \frac{\nu \nu_1}{n!} \nu_1^{n-1} +2D_{\mu\nu} \frac{\nu \nu_1}{n!}\nu^{n-1} + \frac{\nu \nu_1}{n!} (\nu_1^{n-1} -\nu^{n-1}) Y_n^{\mu}]\\
=\ & (\nu-\nu_1)Y_0 + 2D_{\mu \nu_1} \nu (e^{\nu_1} -1) + 2D_{\mu \nu} \nu_1 (e^{\nu} -1) - \nu \nu_1\sum_{n=1}^{\infty} \frac{\nu^{n-1} - \nu_1^{n-1}}{n!} Y_n^{\mu}\\
\leq\ & (\nu-\nu_1)Y_0 + 2D_{\mu \nu_1} \nu (e^{\nu_1} -1) + 2D_{\mu \nu} \nu_1 (e^{\nu} -1)\\
\equiv\ & (\nu-\nu_1)Y_0 + g'(\mu,\nu,\nu_1).
\end{split}
\end{equation}
Note that $\nu^{n-1} - \nu_1^{n-1} \geq 0$ is used above. Thus, from inequality~\labelcref{Y0_bound1}, we can get the lower bound of  $Y_0$
\begin{equation}\label{Y0_bound2}
 Y_0  \geq Y_0^L  = \max \{\frac{1}{\nu-\nu_1} [\nu e^{\nu_1}Q_{\nu_1}-\nu_1 e^{\nu}Q_{\nu}-g'(\mu,\nu,\nu_1)], 0\}.
\end{equation}
Then, we can estimate the lower bound of $Y_1$ as shown below.
\begin{equation}
\label{estimate}
\begin{split}
&e^{\nu}Q_{\nu}-e^{\nu_1}Q_{\nu_1}=\sum_{n=0}^{\infty} \frac{\nu^n}{n!}Y_n^{\nu} -\sum_{n=0}^{\infty} \frac{\nu_1^n}{n!}Y_n^{\nu_1}\\
=\ &Y_0^{\nu}-Y_0^{\nu_1}+\nu Y_1^{\nu}-\nu_1Y_1^{\nu_1} +\sum_{n=2}^{\infty} \frac{1}{n!}(\nu^n Y_n^{\nu} -\nu_1^n Y_n^{\nu_1})\\
\leq\ &\nu(Y_1^\mu +2D_{\mu\nu})-\nu_1(Y_1^\mu-2 D_{\mu\nu_1}) +\sum_{n=2}^{\infty} \frac{1}{n!} [\nu^n (Y_n^\mu+ 2D_{\mu\nu})- \nu_1^n (Y_n^\mu- 2D_{\mu\nu_1})]\\
=\ &(\nu-\nu_1)Y_1^\mu +2 \nu D_{\mu\nu}+ 2\nu_1 D_{\mu\nu_1} +\sum_{n=2}^\infty \frac{1}{n!}[(\nu^n-\nu_1^n)Y_n^\mu+ 2 \nu^n D_{\mu\nu} + 2\nu_1^n D_{\mu\nu_1}]\\
=\ &(\nu-\nu_1) Y_1^\mu +\sum_{n=2}^\infty \frac{1}{n!} (\nu^n-\nu_1^n)Y_n^\mu +2 D_{\mu\nu} (e^{\nu}-1) +2 D_{\mu\nu_1}(e^{\nu_1}-1)\\
\leq\ &(\nu -\nu_1)Y_1^\mu +\frac{\nu^2-\nu_1^2}{\mu^2} (e^{\mu}Q_\mu-Y_0^L-\mu Y_1^\mu) +\frac{\mu(\nu-\nu_1)-(\nu^2-\nu_1^2)}{\mu} g(\mu,\nu,\nu_1)\\
=\ &\frac{\mu(\nu-\nu_1)-(\nu^2-\nu_1^2)}{\mu}Y_1^\mu + \frac{\nu^2-\nu_1^2}{\mu^2}(e^{\mu}Q_{\mu} -Y_0^L) +\frac{\mu(\nu-\nu_1)-(\nu^2-\nu_1^2)}{\mu} g(\mu,\nu,\nu_1).
\end{split}
\end{equation}
\end{widetext}
Here we use the inequality $\frac{\nu^n-\nu_1^n}{\mu^n}\leq \frac{\nu^2-\nu_1^2}{\mu^2}$ for all $n\geq 2$ and $Y_0^L$ is given by~\cref{Y0_bound2}. In~\cref{estimate},
\begin{equation}
g(\mu,\nu,\nu_1)\equiv \frac{2\mu[D_{\mu\nu}  (e^{\nu}-1) + D_{\mu\nu_1}(e^{\nu_1}-1)]}{\mu(\nu-\nu_1)-(\nu^2-\nu_1^2)}.
\end{equation}
We assume $\mu\geq \nu+\nu_1$. By solving inequality~\labelcref{estimate}, the lower bound of $Y_1^\mu$ is given by
\begin{equation}\label{a_y1}
\begin{split}
Y_1^\mu \geq\ &\frac{\mu[e^{\nu}Q_{\nu}\!-\!e^{\nu_1}Q_{\nu_1}\!-\!\frac{\nu^2\!-\!\nu_1^2}{\mu^2}(e^{\mu}Q_{\mu\!}-\!Y_0^L)]}{\mu(\nu-\nu_1)-(\nu^2-\nu_1^2)}  -g(\mu,\nu,\nu_1)\\
\equiv\ &G(\mu,\nu,\nu_1) -  g(\mu,\nu,\nu_1).
\end{split}
\end{equation}
For the weak~+~vacuum decoy state protocol, it is easy to get the lower bound of $Y_0$ from~\cref{Y0_bound2}. That is $Y^L_0= Q_{\nu_1=0}= Q_\text{vac}$. Then~\cref{a_y1} can be written as
\begin{equation}\label{Y1_a}
Y_1^\mu \! \geq\! \frac{\mu}{\mu\nu\!-\!\nu^2} [e^{\nu}Q_{\nu}\!-\!\frac{\nu^2}{\mu^2}e^{\mu}Q_{\mu}\!-\!\frac{\mu^2\!-\!\nu^2}{\mu^2}Q_\text{vac}\!-\!2 D_{\mu\nu}(e^{\nu}\!-\!1)].
\end{equation}

\section{Estimating the upper bound of $e_1^\mu$}
\label{A_e1}
We derive the upper bound of $e_1^\mu$ as follows. Note the fact that
\begin{equation}
e^\omega Q_\omega E_\omega =\sum_{n=0}^\infty \frac{\omega^n}{n!} Y_n^\omega e_n^\omega \geq e_0 Y_0 +\omega Y_1^\omega e_1^\omega,
\end{equation}
Thus, we have
\begin{subequations}
\begin{equation}
e_1^\mu\leq \frac{e^{\mu}Q_\mu E_\mu-e_0 Y_0^L}{\mu Y_1^\mu}\equiv K^{\mu},
\end{equation}
\begin{equation}
e_1^\mu \leq \frac{e_1^\nu Y_1^\nu + 2D_{\mu\nu}}{Y_1^\mu} \leq \frac{e^\nu Q_\nu E_\nu -e_0 Y_0^L +2\nu D_{\mu\nu}}{\nu Y_1^\mu} \equiv K^{\nu},
\end{equation}
\begin{equation}\label{K2}
\begin{split}
e_1^\mu \leq\ & \frac{e_1^{\nu_1} Y_1^{\nu_1} + 2D_{\mu\nu_1}}{Y_1^\mu} \leq \frac{e^{\nu_1} Q_{\nu_1} E_{\nu_1} -e_0 Y_0^L +2\nu_1 D_{\mu\nu_1}}{\nu_1 Y_1^\mu}\\
\equiv\ & K^{\nu_1}.
\end{split}
\end{equation}
\end{subequations}
At the same time, we have
\begin{equation}
\begin{split}
&Q_\mu E_\mu e^\mu -Q_\nu E_\nu e^\nu\\
=\ &\sum_{n=0}^\infty [\frac{\mu^n}{n!} Y_n^\mu e_n^mu-\frac{\mu^n}{n!} Y_n^\mu e_n^mu]\\
=\ &u Y_1^\mu e_1^\mu- \nu Y_1^\nu e_1^\nu +\sum_{n=2}^\infty \frac{\mu^nY_n^\mu e_n^\mu -\nu Y_n^\nu e_n^\nu}{n!}\\
\geq\ &(\mu-\nu)Y_1^\mu e_1^\mu -2D_{\mu\nu}(\nu+ \sum_{n=2}^\infty \frac{\nu^n}{n!}) +\sum_{n=2}^\infty \frac{\mu^n-\nu^n}{n!}Y_n^\mu e_n^\mu \\
=\ &(\mu-\nu)Y_1^\mu e_1^\mu -2D_{\mu\nu}(e^\nu-1).
\end{split}
\end{equation}
Here we use the fact that $Y_0^\mu=Y_0^\nu$ and $e_0^\mu=e_0^\nu$, since there is no difference for the vacuum pulse. Then we have
\begin{equation}\label{K31}
e_1^\mu \leq \frac{e^\mu Q_\mu E_\mu -e^\nu Q_\nu E_\nu +2 D_{\mu \nu}(e^\nu-1)}{(\mu-\nu)Y_1^\mu} \equiv K^{\mu\nu}.
\end{equation}
Similarly, we have
\begin{equation}\label{K32}
e_1^\mu \leq \frac{e^\mu Q_\mu E_\mu -e^{\nu_1} Q_{\nu_1} E_{\nu_1} +2 D_{\mu \nu_1}(e^{\nu_1} -1)}{(\mu-\nu_1)Y_1^\mu} \equiv K^{\mu\nu_1}.
\end{equation}
Therefore, the upper bound of $e_1^\mu$ can be estimated by
\begin{equation}\label{e12a}
e_1^\mu \leq \min\{K^\mu, K^\nu, K^{\nu_1}, K^{\mu\nu}, K^{\mu\nu_1}\}.
\end{equation}

For weak~+~vacuum decoy state protocol, $\nu_1=0$ and $Y_0 = Q_\text{vac}$. Then $Q_{\nu_1} E_{\nu_1} = e_0 Y_0$. Inequality~(\ref{K2}) becomes
\begin{equation}
e_1^\mu \leq \frac{1}{Y_1^\mu},
\end{equation}
which holds by the definition of $e_1$. Also, $K^{\mu \nu_1} = K^{\mu}$. Thus, the upper bound of $e_1^\mu$ can be rewritten as
\begin{equation}\label{e1_a}
e_1^\mu \leq \min\{K^\mu, K^\nu, K^{\mu\nu}\}. 
\end{equation}

\def\bibsection{\medskip\begin{center}\rule{0.5\columnwidth}{.8pt}\end{center}\medskip} 

%

\end{document}